\documentclass{article}

\usepackage[T1]{fontenc}
\usepackage[utf8]{inputenc}
\usepackage[american]{babel}
\usepackage[babel]{csquotes}
\usepackage{microtype}

\usepackage{amsmath, amssymb, amsthm, mathtools}
\usepackage{siunitx}
\usepackage{bm}
\usepackage{textgreek}

\usepackage{float}
\usepackage{booktabs}
\usepackage{longtable}
\usepackage{graphicx}
\usepackage{subfig}

\usepackage{hyperref,cleveref}

\usepackage{acro}
\DeclareAcronym{eb}{short=EB, long=Euler-Bernoulli}
\DeclareAcronym{fem}{short=FEM, long=finite element method}
\DeclareAcronym{ga}{short=GA, long=genetic algorithm}

\newcommand{\dd}{\mathop{}\!\mathrm{d}}
\DeclareMathOperator{\tr}{tr}

\title{The positioning of stress fibers in contractile cells minimizes internal mechanical stress}
\author{Lukas Riedel$^{a,b}$, Valentin Wössner$^{c,d}$, Dominic Kempf$^a$, Falko Ziebert$^{c,d}$, \\
Peter Bastian$^{a\ *}$, Ulrich S. Schwarz$^{a,c,d}$
\footnote{Authors for correspondence: Peter Bastian (peter.bastian@iwr.uni-heidelberg.de) and Ulrich S. Schwarz (schwarz@thphys.uni-heidelberg.de)}\\[2mm]
{\small $^a$Interdisciplinary Center for Scientific Computing, Heidelberg University, Heidelberg, Germany}\\
{\small $^b$Institute for Environmental Decisions, ETH Zürich, Zürich, Switzerland}\\
{\small $^c$Institute for Theoretical Physics, Heidelberg University, Heidelberg, Germany}\\
{\small $^d$BioQuant, Heidelberg University, Heidelberg, Germany}
}

\begin{document}

\maketitle

\begin{abstract}
The mechanics of animal cells is strongly determined by stress fibers, which are
contractile filament bundles that form dynamically in response to extracellular cues. 
Stress fibers allow the cell to adapt its mechanics
to environmental conditions and to protect it from structural damage. 
While the physical description of single stress fibers is well-developed, 
much less is known about their spatial distribution on the level of whole cells.
Here, we combine a finite element method for one-dimensional fibers embedded in
an elastic bulk medium with dynamical rules for stress fiber formation based
on genetic algorithms. We postulate that their 
main goal is to achieve minimal mechanical stress in the bulk material with as few fibers as possible.
The fiber positions and configurations resulting from this optimization task alone are in good agreement with those found in experiments where cells in 3D-scaffolds were mechanically strained at one attachment point.
For optimized configurations, we find that stress fibers
typically run through the cell in a diagonal fashion, similar to reinforcement
strategies used for composite material.
\end{abstract}

\clearpage

\section{Introduction}

The mechanics of animal cells is determined mainly by the actin cytoskeleton, a highly dynamic
network of semiflexible filaments that can form different architectures within the same 
cells \cite{murrell_forcing_2015,banerjee_actin_2020,lappalainen_biochemical_2022}.
For example, during cell migration rapidly polymerizing and branched actin networks are used to push the
cell front forward, while contractile actomyosin networks and
bundles are used to retract the rear of the cell \cite{blanchoin_actin_2014}. Early work with light
and electron microscopy of cultured cells discovered that in mechanically stressful
situations, cells often form clearly visible actin filament bundles called
stress fibers \cite{pellegrin_actin_2007}. They are also 
formed \textit{in vivo} when mechanical stress arises, e.g.\ during 
development, wound healing or in the vasculature \cite{lopez-gay_apical_2020}. In general, it is believed
that their main function is adaptation to mechanically challenging conditions of the environment \cite{burridge_tension_2013,kassianidou_biomechanical_2015,livne_inner_2016}.

The detailed investigation of stress fibers
has revealed that they have a large and diverse range of cellular functions, 
including cell shape control, mechanosensing and wound closure,
and that they come in different flavors that 
differ in cellular location and composition \cite{hotulainen_stress_2006,tojkander_actin_2012}. For example, 
dorsal stress fibers growing out of cell-matrix adhesions at the cell front
have been shown to be composed mainly of actin and its crosslinker $\alpha$-actinin,
while transverse arcs form perpendicular to it and in addition contain
the molecular motor non-muscle myosin II, thus making them contractile \cite{hotulainen_stress_2006}. 
The strongest stress fibers are the highly contractile ventral stress fibers that connect
focal adhesions at the ventral side of a cell. They typically result from the fusion of
dorsal stress fibers and transverse arcs \cite{hotulainen_stress_2006,soine_model-based_2015}.
Peripheral or cortical stress fibers are also contracting between two anchoring focal adhesions
and are responsible for the invaginated shapes of contractile cells adhering at discrete sites of adhesion,
which form by a balance of bulk contractility in the cell and line contractility along the periphery
\cite{bischofs_filamentous_2008,brand_tension_2017}. In contrast to ventral stress fibers,
peripheral stress fibers are not formed by other stress fibers, but through a condensation process from the actomyosin cortex \cite{lehtimaki_generation_2021,vignaud_stress_2021}.
Because they are contractile and directly connected to the extracellular environment, ventral and peripheral stress fibers also
act as mechanosensitive organelles that inform the cells about the
mechanical status of their environment \cite{tojkander_generation_2015}. 
Because stress fibers typically are formed when cells are cultured on relatively stiff substrates
\cite{engler_substrate_2004,prager-khoutorsky_fibroblast_2011}, their mechanosensitivity must
be such that large stresses promote their assembly, in agreement with the overall interpretation
that they have a mechanoprotective function and form on demand in order to shield
cells from excessive mechanical stress, although this never has been proven directly.

Besides filamentous actin, the two main molecular components of stress fibers 
are the actin crosslinker $\alpha$-actinin and 
the molecular motor non-muscle myosin II.
Non-muscle myosin II in stress fibers is organized into 300 nm large minifilaments
that alternate with regions held together by $\alpha$-actinin \cite{hu_long-range_2017}. 
Because of this sarcomeric structure of transverse arcs as well as ventral and peripheral
stress fibers, biophysical models for stress fibers often focused
on their similarities and differences to muscle filaments \cite{dasbiswas_ordering_2018}.
The standard way to probe their mechanical properties is laser cutting,
which typically leads to exponential relaxation curves that can be used to
estimates internal stress and motor activity \cite{kumar_viscoelastic_2006,colombelli_mechanosensing_2009,kassianidou_geometry_2017}.
Similar responses have been reported for stress fibers stretched and compressed on
soft elastic substrates \cite{bernal_actin_2022}.
Stress fibers can also be dissected out of cells and their 
mechanical properties measured with micromanipulators \cite{deguchi_tensile_2006,katoh_isolation_1998}.
Together, these experiments show that stress fibers are well-defined, discrete
organelles that have very different mechanics than the bulk of the cell.

Starting from their similarity with muscle fibers,
mathematical models for stress fibers usually consider a linear array
of contractile elements \cite{besser_coupling_2007,stachowiak_recoil_2009,russell_sarcomere_2009}. 
These sarcomeric units produce force according
to the crossbridge cycle of the myosin II molecular motors. By taking
the continuum limit of many sarcomeric units, a continuum 
theory for stress fibers can be formulated \cite{besser_viscoelastic_2011}.
One-dimensional models for stress fibers can be solved numerically by standard methods such as finite differences
or finite elements, but usually are not scaled up to whole-cell models
due to computational challenges and the underlying molecular complexity.
Recently, the immersed boundary method was used to
study different rheological models for stress fibers embedded into
a bulk material representing the cell \cite{savinov_model_2024}, but no
procedure has been developed before
to predict where they will be positioned
in mechanically challenged cells.

In order to study the effect of the actin cytoskeleton on the level of whole cells, 
often filament models are used in which individual polymers of a certain length 
are considered as one-dimensional beams, which can be bent and stretched. 
Additionally, these filaments might be subject to Brownian dynamics, leading to stochastic partial differential equations, 
which can numerically be studied e.g. with the finite element method (FEM) \cite{cyron_finite-element_2009, lin_combined_2014}. 
The dynamics can then be scaled up to networks, like the actin cortex, consisting of crosslinked and/or contractile filaments. Depending on the amount of deformation, these networks transition from being entropy to bending and finally to being stretching dominated \cite{lin_combined_2014}. Cooperative effects between filaments give rise to nonaffine displacements \cite{heussinger_stiff_2006} and strain hardening \cite{lin_combined_2014}.

In order to include the muscle-like contraction of stress fibers and at the same time
to achieve a whole-cell description, earlier work has used a continuum approach based on 
FEM, that predicted in which directions fibers form in adherent contractile cells  
\cite{deshpande_bio-chemo-mechanical_2006,deshpande_model_2007,ronan_cellular_2013}. 
Muscle-like contractile elements were connected to an anisotropic contractile
material by specifying at each point how strongly the cell contracts in 
different directions. However, due to its continuum nature, this approach 
cannot describe situations that depend on the fact that stress fibers
are discrete and individual entities, in particular the fact that they can 
span large distances, that they can cross
each other or that they can have very different force-generating capabilities \cite{soine_model-based_2015}.

Because the existence of stress fibers is often related to mechanical challenges,
their location typically depends on the geometry of the adhesive environment of the cell, 
which determines how external forces are transmitted onto cells.
While dorsal stress fibers and transverse arcs usually form in the
retrograde flow away from advancing cell membranes, ventral and peripheral stress fibers
are anchored in focal adhesions that are located behind the leading edge, at the sides or at the back. 
Thus, their locations strongly depend on the spreading history of the cell
\cite{thery_cell_2006,kassianidou_extracellular_2019}.
In contrast to these internal stress fibers that run through the 
cytoplasm and are believed to be embedded in a weak network
of actin filaments, peripheral stress fibers lining the cell contour
are closely related to the cell cortex and connect neighboring
adhesions by invaginated arcs \cite{bischofs_filamentous_2008,brand_tension_2017}.
Although being contractile and connected to adhesions like ventral stress
fibers, these peripheral stress fibers might be differently
organized due to their different history \cite{vignaud_stress_2021,lehtimaki_generation_2021}.

\begin{figure}[thp]
    \centering
    \includegraphics[width=\textwidth]{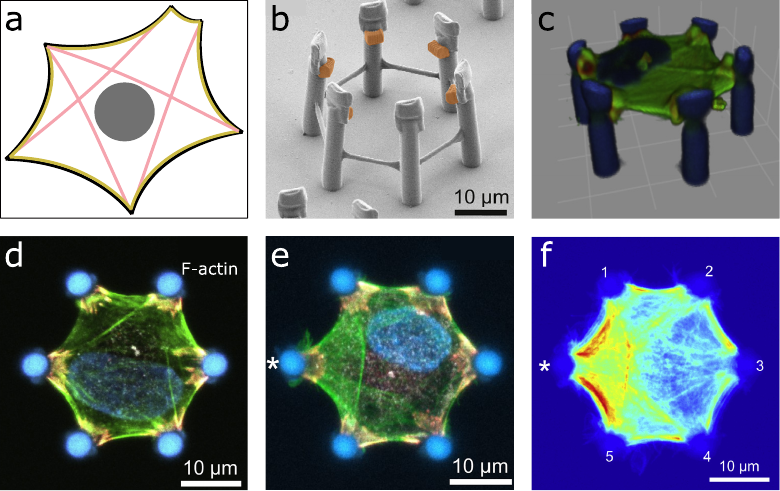}
    \caption{Stress fibers in adherent cells. (a) Cartoon of the stress fiber distribution in a cell adhering to a flat substrate. Peripheral stress fibers (yellow) line the cell contour (black) and in balance with cortical contractility form invaginated arcs. Ventral stress fibers (red) run across the cell in straight lines.
    (b) To study stress fibers without a large external surface, one can use 3D-printed
    scaffolds. This scanning electron micrograph shows a hexagonal arrangement to which cells can adhere
    at the adhesive cubes (orange) \qty{20}{\micro\meter} above ground level.
    (c) 3D reconstruction of a confocal image stack shows that the cells adhere to the scaffold in a flat, almost planar shape with hexagonal symmetry.
    (d) Experimental image of a cell adhering in the scaffold. The F-actin signal (green) and the overall cell shape reflect the six-fold symmetry of the scaffold. Peripheral stress fibers are clearly visible together with a few internal ones.
    (e) Periodic stretching of the left-most pillar (marked with a star) leads
    to reorganization of actin and the formation of new internal stress fibers.
    (f) False color plot shows the actin intensity 
    (with red the highest) averaged over ten cells after \qty{15}{\minute} of periodic stretching of the left pillar with a frequency of \qty{0.5}{\hertz} and a deflection of \qty{5.3}{\micro\meter}.
    All experimental images reproduced with permission from \cite{scheiwe_subcellular_2015}.}
    \label{fig:scheiwe}
\end{figure}

Figure~\ref{fig:scheiwe}(a) summarizes the typical location of ventral and peripheral stress fibers in a schematic drawing
for a cell maturely adhering to a substrate. It is assumed that protruding activity important during spreading has already died down.
Dorsal stress fibers and transverse arcs are excluded from this drawing because they are believed to contribute less to cell mechanics. In order to study \textit{de novo} generation of stress fibers, one can
apply stretching forces to cells, e.g.\ in a cyclic manner \cite{faust_cyclic_2011,greiner_cyclic_2013,livne_cell_2014}.
If in addition one aims at avoiding the effect of surfaces and better
mimic the situation in a 3D tissue, one can culture cells in 
3D-scaffolds \cite{klein_two-component_2011,scheiwe_subcellular_2015,brand_tension_2017,LinkPLOSCB2024}. 
In Figure~\ref{fig:scheiwe}(b), one example of such
a 3D-scaffold is shown as an electron micrograph \cite{scheiwe_subcellular_2015}. The 3D-rendering
in Figure~\ref{fig:scheiwe}(c) demonstrates that indeed
one can obtain cells with the same geometry
as the scaffold, which here is hexagonal. 
Note that in these experiments, the nucleus tends to be positioned in the middle of the
cell, thus excluding a large region of the cells to stress fibers; this is often
different in cells adhering to a flat substrate, where the nucleus typically is positioned
above a layer of ventral stress fibers.
Figure~\ref{fig:scheiwe}(d) shows that in this
configuration, the actin cytoskeleton is
dominated by peripheral stress fibers forming invaginated arcs,
as shown schematically in Figure~\ref{fig:scheiwe}(a) for a cell on a substrate.
If one now oscillates one of the microfabricated adhesion platforms with a
microindenter, internal stress fibers form in response to the external force, 
as shown in Figure~\ref{fig:scheiwe}(e). Although the exact mechanisms for their formation
in 3D are not known, they are believed to resemble ventral stress fibers, because they are 
anchored on both sides to strong adhesions.
The overlay in Figure~\ref{fig:scheiwe}(f) shows
that the newly formed stress fibers 
follow clear geometrical patterns and one can expect that their biological function
would be to protect the cell from mechanical damage, although this is difficult to prove directly.

Here we aim to approach this interesting and important subject from the theoretical side. 
Motivated by the clear experimental results shown in Figure~\ref{fig:scheiwe}, we suggest a
general criterion to predict where such stress fibers
appear in cells that are anchored only at a few adhesion sites, one of which 
is mechanically challenged. In detail, we postulate that stress fibers are generated
in such a way as to minimize mechanical stress in the bulk of the cell.
At the same time, we also consider that this process should not use too much
material, because cytoskeletal resources are limited in cells \cite{goehring2012organelle}.
Our procedures as explained below lead to a quantitative prediction of the location of stress fibers in whole cells,
which is highly relevant both for the basic understanding of cell mechanics
and for practical purposes like predicting cell migration in complex
environments. It also could contribute to the design of synthetic systems
that mimic the mechanical adaptability of living systems, for example
for the design of composite material for airplane wings \cite{benezech2024scalable}. 
In order to upscale to
whole cells, we are inspired by earlier work using FEM-approaches
with anisotropic contractions \cite{deshpande_bio-chemo-mechanical_2006,andersen_cell_2023}.
In contrast to this earlier work, however, here we aim at a FEM-framework 
that describes single discrete stress fibers to which one
can ascribe different properties, reflecting the experimental observation that
each stress fiber can be of a different type \cite{soine_model-based_2015}.

In the following, we introduce a model for whole cell mechanics with stress fibers that are described 
as discrete \ac{eb} beams embedded into a continuous matrix, 
similar to finite element models for fiber-reinforced composites \cite{dodwell_multilevel_2021}. In order
to avoid the large computing times required for explicitly resolved
stress fibers, we use a recently developed finite element method (CutFEM) \cite{hansbo_cut_2017}
that leads to a much reduced computing time. To predict at which positions stress fibers form, 
we consider different possible measures for internal mechanical stress and then 
identify integrated von Mises stress as an appropriate choice. 
We optimize stress fiber distributions using genetic algorithms, because
this method fits well to their discrete and modular structure and also allows
us to implement resource allocation as a side constraint. 
The results of our computer simulations are in very good agreement with 
experiments and thus suggest that the spatial stress fiber configurations found in experiments 
indeed serve the function of reducing intracellular stress.

\section{Methods}

\subsection{Linear elasticity}

In multicellular organisms, it is a major function of single cells to mechanically contribute
to tissue and body shape. Therefore, their effective mechanical laws should contain an elastic
element, as evidenced by the stably invaginated arcs in Figure~\ref{fig:scheiwe}. Indeed, 
whole cell mechanics is often modeled with viscoelastic laws of the
Kelvin-Voigt type \cite{tlili_colloquium_2015,andersen_cell_2023}. Here we focus
on the elastic aspect and model the cell bulk as a 2D linear elastic medium in the plane-stress approximation by assuming a flat cell with negligible out-of-plane stress components. 
In the absence of body force, the problem is then described by the stationary Cauchy momentum equation in two dimensions
\begin{equation}
  \label{eq:linear-elasticity}
  \left[ \nabla_0 \cdot \bm{\sigma}(\bm{u}) \right] = \bm{0},
\end{equation}
for the displacement $\bm{u}$, where $\nabla_0$ denotes the spatial derivative operator with respect to the undeformed reference (Lagrange) configuration.
The Cauchy stress tensor $\bm{\sigma}$ is the sum of a passive component $\bm{\sigma}_p$ that depends on the displacement $\bm{u}$ and an assumed constant active component $\bm{\sigma}_a$,
\begin{equation}
  \bm{\sigma}(\bm{u}) = \bm{\sigma}_p(\bm{u}) + \bm{\sigma}_a.
\end{equation}
The passive component is the usual Cauchy stress tensor for linear elastic materials, defined as
\begin{equation}
  \label{eq:cauchy-stress-tensor}
  \bm{\sigma}_p = 2 \mu \bm{\epsilon} + \lambda \tr(\bm{\epsilon}) \bm{I}.
\end{equation}
Here $\lambda$ and $\mu$ are the 
Lamé constants, $\operatorname{tr}$ denotes the matrix trace and $\bm{I}$ is the identity matrix. The linear strain tensor is given by
\begin{equation}
  \label{eq:strain-tensor}
  \bm{\epsilon} =\frac{1}{2} \left[ \nabla_0 \bm{u} + \left[ \nabla_0 \bm{u} \right]^T \right]
\end{equation}
with the superscript $T$ indicating the matrix or vector transpose.
In plane-stress, the Lamé constants are related to the Young's modulus $E$ and the Poisson ratio $\nu$ via
\begin{equation}
    \lambda = \frac{E \nu}{1 - \nu^2}, \quad \mu = \frac{E}{2(1 + \nu)}.
\end{equation}
Performing all calculations per unit length in the third dimension, both Lamé constants have still the units of stress and are measured in Pascal.
The active stress component $\bm{\sigma}_a$ is the bulk prestress, modeling the myosin motor-induced contraction of the actin network.
For simplicity, we here assume an isotropic prestress that is defined by a single parameter $\varsigma$,
\begin{equation}
  \bm{\sigma}_a = \varsigma \bm{I}.
\end{equation}
A positive value, $\varsigma > 0$, induces a contraction of the medium
and $\varsigma$ can be thought of as being proportional to the motor density and activity.

\subsection{Euler-Bernoulli beam theory}

For the momentum equation of a 1D \acf{eb} beam embedded in a 2D medium, 
we need to decompose the displacement $\bm{u}$ into its normal and tangential components relative to the beam
(in contrast to 3D, for 2D we do not need the binormal).
With $\bm{\hat{n}}$ and $\bm{\hat{t}}$ being the local normal and tangential unit vectors, 
these components are given by
\begin{equation}
  u_n = \bm{\hat{n}} \cdot \bm{u}, \quad u_t = \bm{\hat{t}} \cdot \bm{u}.
\end{equation}
For the fiber we assume a constant Young's modulus $E_f$ and a constant radius $r$. Since we perform calculations per unit length in the third dimension, the effective cross section is given by $2r$.
Its momentum equation in the absence of load or body force then reads
\begin{equation}
  \label{eq:euler-bernoulli-momentum}
  E_f \frac{(2r)^3}{12} \partial_{tt}^2 \partial_{tt}^2 u_n + 2 E_f r \partial_{tt}^2 u_t = 0,
\end{equation}
where $\partial_t$ denotes the spatial derivative in tangential direction with definitions
\begin{equation}
  \partial_t = \bm{\hat{t}} \cdot \nabla_0, \quad \partial_{tt}^2 = \partial_t \partial_t.
\end{equation}
The first term in Equation~\eqref{eq:euler-bernoulli-momentum}
describes bending of the beam and the second term contraction/extension. 

Because the \ac{eb} beams serve to represent stress fibers, 
we have to account for a contractile prestress in tangential direction,
due to the action of the myosin motors. 
Similarly to the bulk medium, 
we assume a prestress parameter $\varsigma_f$ for fibers, implying
\begin{equation}
  \sigma_a = 2 r \varsigma_f.
\end{equation}
Homogeneous contraction should not induce bending, hence
the active component only enters the stress contribution in the second term of Equation~\eqref{eq:euler-bernoulli-momentum}, resulting in
\begin{equation}
  \label{eq:euler-bernoulli-momentum-prestress}
  E_f \frac{2r^3}{3} \partial_{tt}^2 \partial_{tt}^2 u_n + 2 r \partial_t \left[ E_f \partial_t u_t +  \varsigma_f \right] = 0.
\end{equation}
Again, positive $\varsigma_f$ corresponds to a contraction of the fiber.

\subsection{Finite element method}

\begin{figure}%
    \centering
    \includegraphics[width=\textwidth]{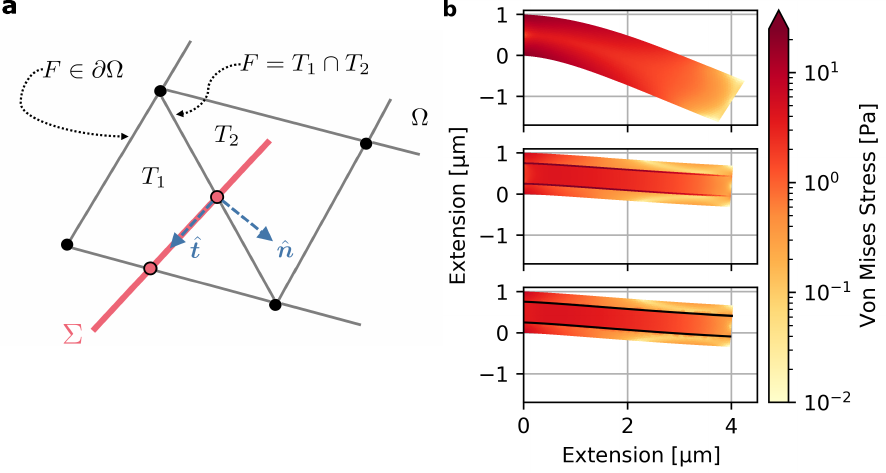}
    \caption{Embedded beam approach using CutFEM. (a) Spatial discretization of the finite element method with embedded beams. $\Omega \in \mathbb{R}^2$ indicates the modeled two-dimensional elastic domain that is tesselated into grid cells $T_i$.
    The one-dimensional \ac{eb} beam, given by $\Sigma$, is a straight line in the displayed reference configuration with normal vector $\hat{\bm{n}}$ and tangential vector $\hat{\bm{t}}$.
    (b) Bending a composite material under own weight as an example: the structure, with Young's modulus of \qty{100}{\pascal} and Poisson ratio of 1/3, is fixed at the left side and loaded with a body force of
    \qty{-0.4}{\pascal\per\micro\meter}
    in vertical direction. Top: no \ac{eb} beams. Middle: two resolved \ac{eb} beams with a Young's modulus 100-fold larger than the Young's modulus of the bulk, same Poisson ratio and a diameter of \qty{0.05}{\micro\meter}. Bottom: two embedded \ac{eb} beams, indicated by black lines, with the same properties as the resolved ones. The resulting deformation of the structure in the embedded description is the same as in the resolved one.}
    \label{fig:discretization}
\end{figure}

In a first, so-called ``resolved'' modelling approach, stress fibers are modelled as elastic cylinders with
their own Young's modulus, Poisson ratio and prestress embedded in a bulk medium. In the standard finite
element method for linear elasticity, see e.g. \cite{braess_2007}, the computational domain is partitioned into mesh elements
and the displacement is approximated by a continuous and element-wise polynomial function. An
accurate approximation of the stress fibers requires the mesh to resolve them and to be sufficiently fine.
Since stress fibers are long, thin objects with a diameter less than one hundredth of the cell diameter, 
this will lead to a challenging meshing problem, in particular in three spatial dimensions. 
Moreover, the number and exact location of stress fibers develops dynamically and requires
repeated remeshing.

In a second, so-called ``embedded beam'' modelling approach, the complex 
(re-)meshing problem is avoided by employing a so-called cut finite
element method \cite{hansbo_2004,engwer_2009,hansbo_cut_2017}.
Cut-cell methods construct finite element functions on a (simpler) background mesh and then restrict
their support to the intersection of the background mesh with the complex domain. A further
reduction of complexity can be achieved by representing the stress fibers as one-dimensional
straight-line objects of constant thickness in the undeformed domain. 
In Reference~\cite{hansbo_cut_2017}, a cut-cell finite element scheme
for one-dimensional EB beams embedded in a two-dimensional linear elastic material was developed.
We extended this scheme by a prestress term and employ it to model stress fiber systems.
The starting point of the scheme in \cite{hansbo_cut_2017} is a standard conforming quadratic finite 
element method on triangles for two-dimensional linear elasticity equations in displacement formulation of Equation~\eqref{eq:linear-elasticity}.
The corresponding conforming triangular mesh constitutes the background mesh for a cut-cell approximation
of the fibers, i.e. the mesh is intersected with the straight-line
fibers in the undeformed configuration, cf.~Figure~\ref{fig:discretization}(a), leading to a piecewise quadratic function on line
segments for the EB beam formulation. Since Equation~\eqref{eq:euler-bernoulli-momentum-prestress} is a fourth-order equation in the normal
derivatives, a $C^0$
interior penalty discontinuous Galerkin approach \cite{brenner_2017} is used to avoid the need for continuity of
normal derivatives in the finite element ansatz. As the linear elasticity FEM for the bulk medium and the $C^0$ interior penalty DG method require the exact same solution spaces, their coupling can be implemented by adding the stiffness matrices of both problems. For the exact formulation of the method we refer to \cite{hansbo_cut_2017}. 

The finite element scheme has been implemented in the DUNE software framework \cite{dune_2008,dune_2021}
using its automatic code generation facility \cite{kempf_2020} to reduce coding effort of the
cut-cell scheme significantly. 
The arising linear systems of equations are solved with the sparse direct solver UMFPACK \cite{UMFPack_2004}. 
Following an example described in \cite{hansbo_cut_2017},
Figure~\ref{fig:discretization}(b) demonstrates the effect of fibers reinforcing a structure, which is fixed at one side and bends under its own weight. The Youngs's modulus of the fibers is 100-fold larger than the one of the bulk. The displacement and bulk von Mises stress (defined in Equation~\eqref{eq:von-Mises-stress}) of the embedded approach agree very well with the resolved description. Before using the embedded beam approach to evaluate complex
configurations with many stress fibers, 
below we will compare the resolved and embedded beam approaches, including prestress and localized external load, to ensure that both approaches produce similar results when representing stress fibers.

\subsection{Scalar observables for mechanics}

As discussed in the introduction,
to find locations where stress fibers 
are generated by the cell,
we do not resort to any detailed
biophysical model. Rather, we formalize the assumed universal purpose of all stress fibers: they should form to oppose external forces and
thus protect the bulk material.
Here we propose two scalar quantities to assess
the effect of the stress fibers.
First, a practical and often used scalar for quantifying stress in a material
is the von Mises stress. In the plane-stress approximation, it is given by
\begin{equation} \label{eq:von-Mises-stress}
    \sigma_\text{vM} = \sqrt{\sigma_{p,xx}^2 - \sigma_{p,xx} \sigma_{p,yy} + \sigma_{p,yy}^2 + 3 \sigma_{p,xy}^2},
\end{equation}
with the components $\sigma_{p,ij}$ of the Cauchy stress tensor, given by Equation~\eqref{eq:cauchy-stress-tensor}. 
We do not include the active bulk stress here
because it is also a cause of deformations and
we aim to minimize their effect. 
This implies that
we consider the undeformed state prior to any deformation due to bulk contraction or external
perturbations as our reference state. 
Because we focus on the passive part of the
stress, a similar scalar measure would be 
the linear strain energy density as expressed
via the strain tensor
\begin{equation} \label{eq:strain energy}
    \mathcal{W} = \frac{1}{2} \lambda [\epsilon_{kk}]^2 + \mu \left[ \epsilon_{ij} \epsilon_{ij} \right],
\end{equation}
where we use the sum convention for repeated indices.
Both quantities are defined at every point of the elastic medium.
Note that for both measures we only consider the bulk contributions and do not include the stress or strain within the fibers. In this way, the values
of our chosen measures become a function of
the fiber characteristics, allowing for an assessment of the protection provided by a fiber configuration of interest.

To define a global loss function, we calculate cell-averaged quantities by integrating over the deformed cell volume $\Omega$ 
akin to an $L^1$-norm,
\begin{equation}
    \lVert \cdot \rVert = \int_\Omega \lvert \cdot \rvert \dd V. 
\end{equation}
The task of our optimization algorithm is now to minimize the global mechanical observables $\lVert \sigma_\text{vM} \rVert$ and $\lVert \mathcal{W} \rVert$ inside the cell in response to external forces. Note that both quantities have the units of forces instead of energies, since we integrate over a 2D volume.

\subsection{Genetic algorithm for optimization}
\label{sec:genetic-algorithm}

Genetic algorithms are used for multi-objective optimization and require only little adaptation to the specific optimization problem at hand \cite{konak_multi-objective_2006}.
We here employ a variation of the elitist, nondominated sorting genetic algorithm II (NSGA-II) by \cite{deb_fast_2002}.
Genetic algorithms aim to minimize a set of $K$ objective functions,
\begin{equation}
    \bm{z}(\bm{x}) = \left\lbrace z_1(\bm{x}), \ldots, z_K(\bm{x}) \right\rbrace,
\end{equation}
with respect to a decision variable set, $\bm{x}$.
As the objective functions are usually conflicting, the algorithms optimize according to a domination measure, where
\begin{align}
     \bm{x} \prec \bm{y}, \, \,\,\text{if}\,\ \, &z_i(\bm{x}) \leq z_i(\bm{y}) \, \forall i \in \{1, \ldots, K\} \,\, \text{and}\\
        &z_j(\bm{x}) < z_j(\bm{y}) \,\, \text{for at least one} \, j \in \{1, \ldots, K\},
\end{align}
in which case we say that ``$\bm{x}$ dominates $\bm{y}$.''
The set of decision variables that are not dominated by others are considered \emph{Pareto optimal} and form the so-called \emph{Pareto front} in the objective function space. 
All these configurations are considered as optimal solutions to the problem. 
In our specific case,
a population of decision variable sets 
corresponds to a certain stress fiber configuration and the objective function
can be cell-averaged stress, or elastic energy,
and the amount of fiber material.

Each iteration of a genetic algorithm then updates a population of decision variable sets to better resemble the true Pareto front.
It applies an update according to a pseudo-evolutionary dynamics, 
which consists of three steps:
(i) \emph{selection} of decision variables whose objective function values are considered closest to the optimal Pareto front,
(ii) \emph{crossover} of elements from selected decision variable sets to create ``offspring'' sets, and
(iii) \emph{mutation} of decision variables in the offspring sets.
An elitist algorithm preserves variable sets from previous iterations by performing selection simultaneously on the ``parent'' and on the ``offspring'' population in each iteration.
In the following, we will elucidate the algorithm steps in more detail.

\paragraph{Selection}
Let $P_t$ denote the parent population and $Q_t$ denote the offspring population of decision variable sets at algorithm iteration $t \in \mathbb{N}^+$.
The selection of an elitist genetic algorithm operates on the total population $R_t = P_t \cup Q_t$.
NSGA-II performs selection by choosing decision variables according to non-dominated fronts and placing them into the new parent population $P_{t+1}$ until the original parent population size is reached, $|P_{t+1}| \geq |P_t|$, where $\lvert\cdot\rvert$ denotes the set cardinality.
A non-dominated front of the full population $R_t$ is defined as the set of non-dominated decision variable sets,
\begin{equation}
    F_i(R_t) = \left\lbrace \bm{x} \in R_t : \bm{x} \not\succ \bm{y} \, \forall \bm{y} \in R_t \setminus \bigcup_{j=1}^{i-1} F_j \right\rbrace,
\end{equation}
and the new parent population is the union of these fronts until it reaches the size of the previous parent population,
\begin{equation}
    P_{t+1} = \left\lbrace F_1(R_t), \ldots, F_L(R_t) : \left\lvert \bigcup_{i=1}^{L-1} F_i(R_t) \right\rvert < \lvert P_t \rvert \leq \left\lvert \bigcup_{i=1}^{L} F_i(R_t) \right\rvert \right\rbrace,
\end{equation}
where $L$ indicates the number of non-dominated fronts selected.
The last selected front typically makes $P_{t+1}$ larger than $P_t$.
To ensure a steady size of the parent population, we order the decision variables from the last non-dominated front $F_L$ by crowding distance, a measure of distance between three adjacent decision variables in objective function space.
We then drop the decision variables with the lowest crowding distance score from the new parent population, until its size matches the one of the previous parent population \cite{deb_fast_2002}.

\paragraph{Crossover}
To create the new offspring population $Q_{t+1}$, we repeatedly perform single-point crossover to create two offspring variable sets from two randomly chosen parent variable sets out of $P_{t+1}$.
The relative probability of a variable set $\bm{x}$ to be selected as parent is given by its fitness
\begin{equation}
    f(\bm{x}; t+1) = \left[\varrho(\bm{x}; t+1)\right]^{-1},
\end{equation}
defined as the inverse of the variable set rank
\begin{equation}
    \varrho(\bm{x}; t+1) = 1 + \sum_{\bm{y} \in P_{t+1}, \bm{y} \prec \bm{x}} \varrho(\bm{y}),
\end{equation}
which is one plus the sum of ranks of all variable sets that dominate $\bm{x}$.
This ensures that dominated decision variables receive an escalating rank, implying a lower fitness, and hence a strongly reduced probability of being selected for crossover.
Single-point crossover between two selected variable sets $\bm{x}, \bm{y} \in P_{t+1}$ then chooses a random crossover index $i_c \in \{ 2, \ldots, \min(|\bm{x}|, |\bm{y}|) - 1\}$, and creates the offspring
\begin{align}
    \bm{x}' &= \left( x_1, \ldots, x_{i_c}, y_{i_c + 1}, \ldots, y_{|\bm{y}|} \right),\\
    \bm{y}' &= \left( y_1, \ldots, y_{i_c}, x_{i_c + 1}, \ldots, x_{|\bm{x}|} \right),
\end{align}
which become part of the new offspring population $Q_{t+1}$.

\paragraph{Mutation}
Variable sets in the new offspring population $Q_{t+1}$ are then mutated to introduce variation in the population.
Mutation of a variable set may include the modification of single or multiple variables in the set, the addition of variables, and the deletion of variables.
The exact mechanism strongly depends on the type of optimization problem and the information encoded in the variable sets. Details on the implementation for our specific application to stress fiber configurations are given below.

\section{Results}

\subsection{Effect of stress fibers on internal cell stress}

\begin{figure}[t!]
    \centering
    \includegraphics[width=\textwidth]{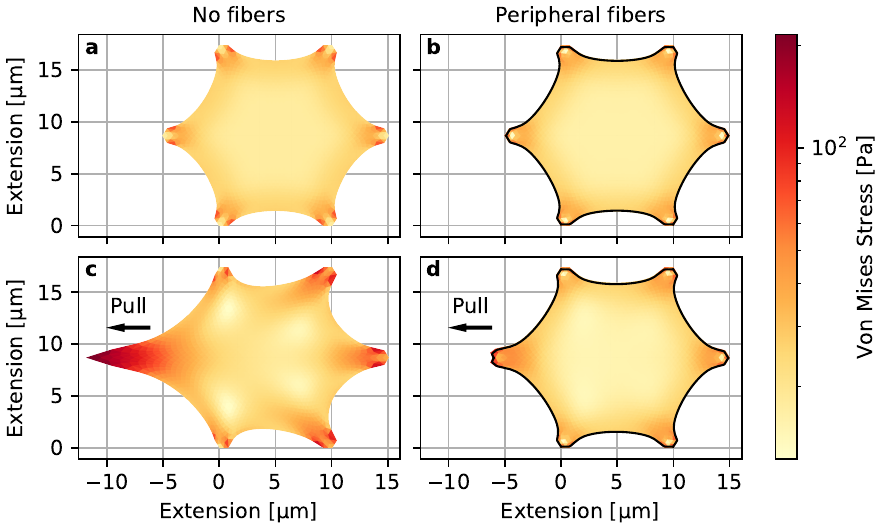}
    \caption{Contractile cell without and with peripheral stress fibers.
    (a) Simulation of a contractile cell with six adhesion points. We have bulk stiffness of \qty{100}{\pascal} and active stress of \qty{30}{\pascal}, but no stress fibers. (b) Same situation, but now with peripheral stress fibers. The \acl{eb} beams have a stiffness of \qty{10}{\kilo\pascal}, a tangential prestress of \qty{1.2}{\kilo\pascal} and a radius of \qty{50}{\nano\meter}.
    (c) Without stress fibers, a force on the leftmost adhesion leads to large von Mises stress in the cell (logarithmic color code). (d) With stress fibers, these large stresses do not develop and the overall displacement is much reduced.}
    \label{fig:scheiwe-model}
\end{figure}

In Figure~\ref{fig:scheiwe-model} we present 
results for simulations of a contractile cell without and
with peripheral stress fibers, using the embedded EBs. 
In general, we select a hexagonal cell with a circumradius of \qty{10}{\micro\meter} 
as our reference cell configuration. This shape has the advantage that it is sufficiently
detailed to allow the system to develop 
different responses, but at the same time, its regular
geometry gives clear results. 
The corners are assumed to be strongly pinned by
focal adhesions, thus
we employ a Dirichlet boundary condition on the displacement field there. In contrast, the cell edges are free. 
This cell is then allowed to contract
under its self-generated homogeneous stress $\bm{\sigma}_a$.
As seen in Figure~\ref{fig:scheiwe-model}(a), 
invaginated arcs form between neighboring adhesions.
The presence of peripheral stress fibers in
Figure~\ref{fig:scheiwe-model}(b) does not make a big
difference, although a detailed analysis shows that 
these stress fibers are stretched by 7 percent and that
they slightly change the cell shape. In Figure~\ref{fig:scheiwe-model}(c),
we now pull at the leftmost corner. This is achieved by adding a 
boundary force $\tau(y)$ using a Neumann boundary condition depending on the vertical extension $y$.
The force has a maximum of \qty{250}{\pascal} directly at the corner and decreases in y-direction in a Gaussian shape with a standard deviation of \qty{0.5}{\micro\meter}.
One can clearly see that the cell without stress fibers is stretched severely
under the action of the force.
When now simulating with peripheral stress fibers
in Figure~\ref{fig:scheiwe-model}(d), 
the internal cell stress is dramatically
reduced, demonstrating that stress fibers
indeed have a strong role in protecting cells
from mechanical challenges.

\subsection{Comparing embedded and resolved fibers}
\label{sec:fiber-compare}

\begin{figure}[t!]
    \centering
    \includegraphics[width=\textwidth]{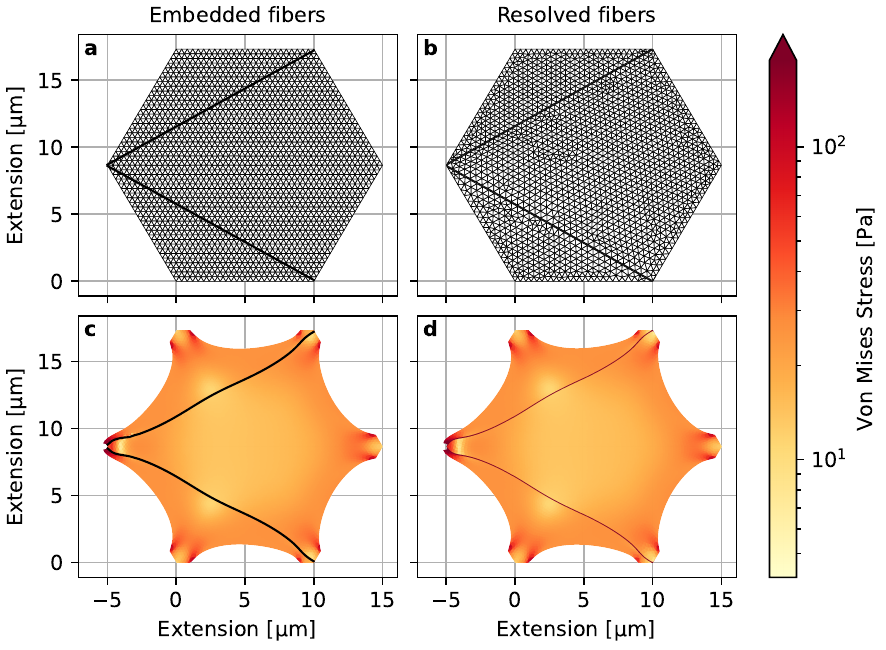}
    \caption{Comparison between embedded and resolved fibers.
    (a) The grid for embedded fibers does not need any adaptation to the stress fiber (visualized as dark lines, radius not to scale). (b) For resolved fibers, the grid has to be remeshed around the fiber. 
    (c) Simulation of force application to the left corner with embedded fibers, with the von Mises stress color-coded in logarithmic scale. Note that the \ac{eb} fiber model does not represent the stress inside the fibers. (d) The resolved fiber gives the same result. Now the stress inside the fiber is directly accessible and it is 
    more than one order of magnitude larger than within the bulk of the cell.}
    \label{fig:comparison}
\end{figure}

In general, stress fibers could occur at any position in the cell and thus we have to check that the CutFEM method also works for more complicated situations.
Therefore we next addressed the question of the difference between
simulating embedded versus resolved fibers for more complex situations. 
To that end, we  used the same hexagonal cell setup as just discussed.
As before, five cell corners were fixed
and at the left a force was applied.
In Figure~\ref{fig:comparison},
we include, as an illustrative configuration, two internal 
stress fibers with a radius of \qty{50}{\nano\meter} oriented ``diagonally'' 
and hence opposing the pulling force towards the left side of the cell.
For the \ac{eb} model, the fibers need not be represented in the grid.
Therefore, we create a grid with a Delaunay triangulation only using the hexagonal outline and a target mesh constant of \qty{400}{\nano\meter}, which we have already used in Figure~\ref{fig:scheiwe-model}.
In contrast, for the explicit representation of the fibers inside the 2D elastic medium, we need to include the fiber geometry into the grid and rerun the triangulation, resulting in a grid with more nodes even though the mesh constant is the same.
The resulting grids are displayed 
in Figure~\ref{fig:comparison}(a) and (b), respectively.
For the simulations both grids were refined once by subdividing each triangle 
into four congruent smaller triangles in order to obtain more accurate results and allow for better comparison.

The simulation results shown in Figure~\ref{fig:comparison}(c) and (d) used the same parameters as above
in Figure~\ref{fig:scheiwe-model}
and are again color-coded by the von Mises stress.
We find that both the cell deformation and the stress distribution in the cell bulk medium 
calculated by the two methods match almost perfectly. At the left corner, where the force is applied, a subtle difference is notable. The external force acts on several boundary nodes belonging to the explicit fibers, while the contributions of the implicit fibers, in this configuration, are assigned to a single boundary node, which in turn only indirectly ``protects'' the surrounding ones. This discrepancy becomes smaller the finer the grids are. Note that, in general, small deviations are actually expected, because the explicit fibers are still modelled as an isotropic material, where active contraction occurs also perpendicular to its orientation. That is not the case in the implicit \ac{eb} model, since we only consider an active prestress in tangential direction, cf.~Equation~\eqref{eq:euler-bernoulli-momentum-prestress}.

The region between the fibers right behind the pulled corner is shielded by the two fibers and experiences almost no von Mises stress. The largest stresses appear directly at the pulled edge and at the other corners. 
The stress inside the fibers in (d) is much higher than in the surrounding medium, ranging up to \qty{1}{\kilo\pascal}, and is therefore not represented in the color-code.
The \ac{eb} fiber model considers these stresses as well, but does not explicitly quantify them. In principle, these stresses are accessible in post-processing. However, since we are mainly interested in the stress and strain of the bulk, we do not compute and display the stresses inside the fibers here.

The overall agreement of this comparison demonstrates the validity of the \ac{eb} model also in the presence of an active prestress and in more complex geometries, extending the analysis of the cut cell FEM method presented in reference \cite{hansbo_cut_2017}. In the context
of our work, it is a clear advantage that no
remeshing will be necessary as the stress fiber
distribution changes.

\subsection{Effect of changing stress fibers}
\label{sec:fiber-growth}

\begin{figure}[t!]
    \centering
    \includegraphics[width=\textwidth]{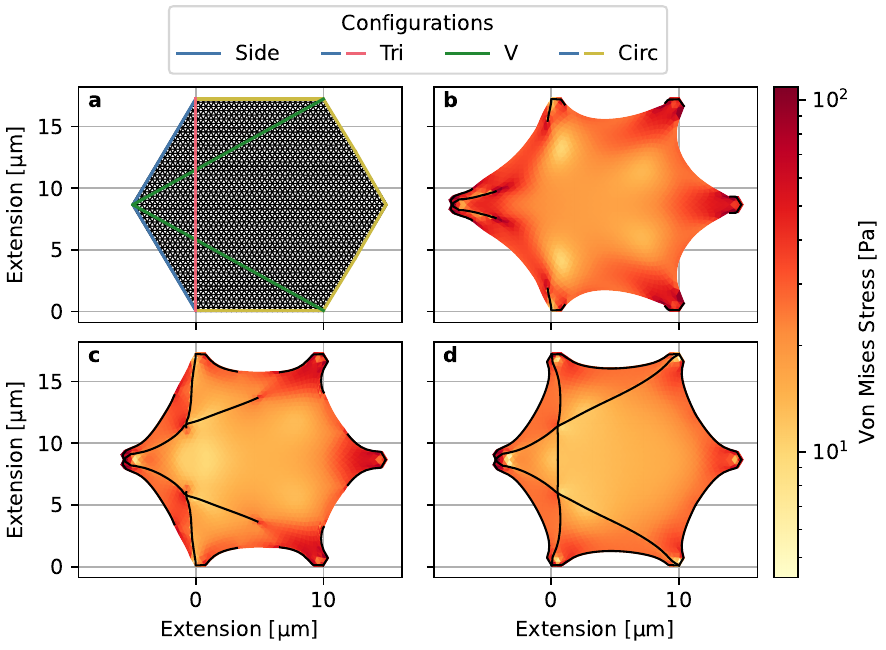}
    \caption{Effect of different fiber types and lengths. (a) We consider four different fiber configurations based on experimental observations. (b)-(d) Simulation results for quasi-static growth at different fiber lengths of (b) 30 $\%$, (c) 60 $\%$ and (d) 100 $\%$ of the final length. The von Mises stress is represented by the color-code in logarithmic scale.}
    \label{fig:growing}
\end{figure}

\begin{figure}[t!]
    \centering
    \includegraphics[width=0.733\textwidth]{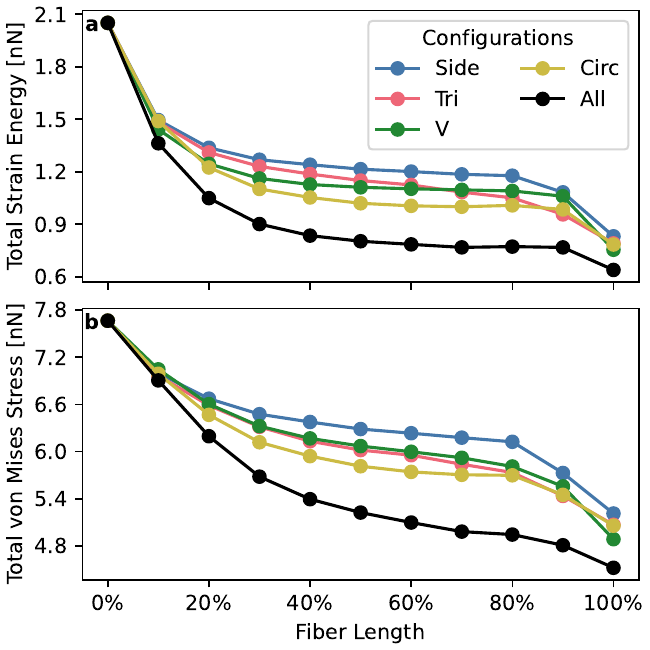}
    \caption{Scalar measures for internal cell stress. (a) Integrated strain energy and (b) integrated von Mises stress as functions of the stress fiber extension for the different fiber configurations.}
    \label{fig:extending}
\end{figure}

Next we addressed the question how changing 
positioning and length of stress fibers 
affects the stress distribution inside cells.
These investigations are also helpful to 
decide which optimization principle is most
appropriate in our context. We start by
defining different types of stress fibers
in our implementation, motivated also by
the experimental observations of stress fibers
described in the introduction. 
Figure~\ref{fig:growing}(a) displays these 
distinct configurations, labeled as follows:
\begin{itemize}
    \item \emph{Side:} The two fibers (blue in Figure~\ref{fig:growing}(a)) in direct contact to where the force is applied
    \item \emph{Tri:} The fibers of \emph{Side} together with a vertical fiber (red) connecting the former
    \item \emph{V:} The two diagonal fibers (green),
    already studied in Figure~\ref{fig:comparison},
    connecting the point of force with the right corners.
    \item \emph{Circ:} All peripheral fibers,
    as already studied in Figure~\ref{fig:scheiwe-model},
    including the ones of \emph{Side}
    \item \emph{All:} All fibers from the four configurations above.
\end{itemize}
Since we are only interested in the steady state 
distributions of stress fibers and not in their
dynamics, we simulated fiber growth by computing stationary solutions for configurations with fiber segments of varying length, chosen from the families above.
Except for the fibers of the \emph{Side} and \emph{V} configurations, where the force 
is directly applied at one of the starting points of the fiber, we let fibers grow from both attachment points symmetrically and meet at the center, see Figure~\ref{fig:growing}(b)-(d).
The material properties and boundary conditions were chosen as in the previous section.

The effect of the fiber growth on the mechanics
was investigated by calculating the integrated
von Mises stress $\lVert \sigma_\text{vM} \rVert$ and the integrated energy density $\lVert \mathcal{W} \rVert$, 
as displayed in Figure~\ref{fig:extending}.
This shows that 
both quantities are maximal when there are no stress fibers in the cell.
As the stress fibers grow, the von Mises stress monotonously decreases and reaches a boundary minimum for fully elongated stress fibers for all fiber configurations. As expected, all four configurations together (\emph{All}) result in the lowest stress for all fiber lengths.
The \emph{Circ} configuration has the largest contribution in stress reduction for not fully extended fibers, while at the end the \emph{V} configuration with only two fibers outperforms \emph{Circ}. In addition, \emph{Side}, as a part of \emph{Circ}, almost leads to the same final von Mises stress as \emph{Circ}, indicating that the fibers close to the pulled corner are the most important and, as already observed in Figure~\ref{fig:scheiwe-model}, stretching is more relevant than bending. 

Interestingly, in contrast to this simple behavior of the von Mises stress, the strain energy remains,
after an initial drop, mostly constant for increasing fiber length and is only dropping again when fibers are fully extended.
Notably, the energy minimum may not be reached for a given fiber configuration, which could lead to suboptimal results in the optimization process.
These observations suggest
that the von Mises stress is a more suitable
objective function due to its monotonous behavior and the better performance of fibers, which directly oppose the external force.

\subsection{An optimization scheme for fiber distribution}
\label{sec:fiber-cfgs}

We now define an optimization scheme for finding stress fiber configurations in a cell based only on mechanical parameters and solve it using a genetic algorithm as introduced in section \ref{sec:genetic-algorithm}.
Building on the simulation setup discussed
in the previous section, we choose the placement and radii of the stress fibers as decision variables.
The component of a decision variable $x_i \in \bm{x}$ is then a set of two coordinates $\bm{p}_i^{(1)}, \bm{p}_i^{(2)}$, which give the start and end position of the fiber $i$, respectively, as well as the fiber's radius $r_i$,
\begin{equation}
    x_i = \left\lbrace \bm{p}_i^{(1)}, \bm{p}_i^{(2)}, r_i \right\rbrace.
\end{equation}
The decision variable then is a $N$-tuple of these,
\begin{equation}
    \bm{x} = \left( x_1, \ldots, x_N \right), \quad N \in \left\lbrace N_\text{min}, \ldots, N_\text{max} \right\rbrace.
\end{equation}
The total number of stress fibers in a cell varies, both between cell types and the given situation. We choose the minimum number of stress fibers as $N_\text{min} = 3$ and the maximum as $N_\text{max} = 30$, in addition to the six peripheral fibers called \emph{Circ} in the previous section. 

As discussed in the introduction and investigated for the specific fiber positions in the last section, we assume that the stress fiber configuration aims to minimize the overall stress in the cell, while in addition using as little material as possible.
Ignoring the underlying biochemical signalling and assembly processes, this suggests two objective functions.
First, the total von Mises stress,
\begin{equation}
    \label{eq:objective-func-stress}
    z_1(\bm{x}) \coloneqq \lVert \sigma_\text{vM} \rVert,
\end{equation}
which technically is not a function of $\bm{x}$, but computed from the elasticity model of the cell including the fibers encoded by $\bm{x}$.
And second, the total volume, here in two dimensions the total area, respectively, of the stress fibers,
\begin{equation}
    \label{eq:objective-func-volume}
    z_2(\bm{x}) \coloneqq \sum_{i=1}^N 2 r_i \left\lVert \bm{p}_i^{(1)} - \bm{p}_i^{(2)} \right\rVert_2,
\end{equation}
where $\lVert \cdot \rVert_2$ indicates the usual Euclidean vector norm.

We chose a parent and offspring population size of 200, $\lvert P_t \rvert = \lvert Q_t \rvert = 200$.
The initial population $Q_1$ consists just of random fiber configurations. Specifically, it is
created by choosing a random number of fibers $N$ from a uniform distribution, $N \sim \mathcal{U}(N_\text{min}, N_\text{max})$, for each decision variable, and then selecting two random, uniformly distributed points inside the 2D grid and a random radius from a normal distribution, $R \sim \mathcal{N}(\mu = \qty{50}{\nano\meter}, \sigma = \qty{15}{\nano\meter})$, for each fiber.
The two points define a line which is extended until the start and end points, $\bm{p}^{(1)}$ and $\bm{p}^{(2)}$, are close to the grid boundary (i.e.~the cell membrane).

Selection is trivial in the first iteration, as all initial population members are selected, $P_1 = \emptyset, P_2 = Q_1$. However, the process is still required to determine the rank and hence fitness for each initial decision variable in order to subsequently apply crossover.
After selection of the parent population $P_t$, we apply crossover to create an offspring population $Q_t$, as presented in section~\ref{sec:genetic-algorithm}.
Crossover is stopped when the offspring population size equals \qty{90}{\percent} of the parent population size.
The remainder of the offspring population is filled up again by random fiber configurations as outlined above.
We introduce this procedure to the NSGA-II algorithm in order to ensure a sufficient randomness of not only single fibers but also entire fiber configurations.

The offspring population is then mutated by iterating over every decision variable $\bm{x} \in Q_t$ and performing the following alteration steps:
\begin{enumerate}
    \item \emph{Removal} of a randomly chosen fiber with probability $p_r = 0.7$, if $|\bm{x}| > N_\text{min}$.
    \item \emph{Variation} of fibers with probability $p_v = 0.7$ for each fiber, by drawing new fiber coordinates and radii from normal distributions whose mean is given by the respective current value.
    This effectively implements a random walk of the single fiber parameters,
    \begin{align}
        \bm{p}_i^{(j)}{}' &\sim \mathcal{N} \left( \bm{\mu} = \bm{p}_i^{(j)}, \bm{\Sigma} = \left[ \qty{400}{\nano\meter} \right]^2 \bm{I} \right),\\
        r_i' &\sim \mathcal{N}(\mu = r_i, \sigma = \qty{15}{\nano\meter}),
    \end{align}
    where the dashed quantities indicate the parameters after variation and $\mathcal{N}(\bm{\mu}, \bm{\Sigma})$ is a multivariate normal distribution with mean $\bm{\mu}$ and covariance matrix $\bm{\Sigma}$, $\bm{I}$ the identity.
    \item \emph{Addition} of one fiber with probability $p_a = 0.7$, if $|\bm{x}| < N_\text{max}$.
    With an additional probability of $p_s = 0.8$, the placement of the new fiber is not random but 
    chosen according to local maxima in the von Mises stress distribution.
    For this, one local maximum is selected with a relative probability of $p = \sqrt{\sigma_\text{vM}}$, with $\sigma_\text{vM}$ the von Mises stress of the local maximum.
    The position of the selected maximum is used as center point of the new fiber, and the direction of the fiber is determined along the direction of the principal stress component at this location.
    Again, the fiber is then extended in both directions until the start and end points lie sufficiently close to the grid boundary.
\end{enumerate}
After mutation, the entries of each decision variable are shuffled.

Finally, the fiber configuration defined by each decision variable of the new offspring population is inserted into the elasticity model and the first objective function, i.e.~the overall stress Equation~\eqref{eq:objective-func-stress}, is calculated. 
The second objective function, the total stress fiber volume, Equation~\eqref{eq:objective-func-volume}, is computed directly from the decision variables.
Because we focus only on internal fibers, which 
form during mechanical challenges, the peripheral fibers are excluded from the genetic algorithm and do not contribute to the computation of the total fiber volume through objective function $z_2$. 

As a model refinement, we also investigate the effect of the cell's nucleus on the stress fiber configurations computed by the genetic algorithm. The nucleus is implemented as an elastic material in the center of the grid with different material properties.
In the reference configuration, 
the nucleus is round with a radius of \qty{3}{\micro\meter}.
The reference grid was adjusted such that grid element interfaces coincide with the nucleus outline.
The nucleus bulk material is chosen to be ten times stiffer than the surrounding cytoplasm \cite{caille_contribution_2002, wohlrab_mechanical_2024}, with a Young's modulus of $E = \qty{1}{\kilo\pascal}$ and
Poisson ratio $1/3$. 
Since the contractile stress originates from the cytoskeleton, which is excluded from the nucleus, we assume the isotropic prestress to be zero within the nucleus. 
The fiber placement by the genetic algorithm has now to be slightly adapted to avoid fibers passing through the nucleus.
To that end, the stress-based placement iteratively selects local stress maxima based on their relative probability, as outlined above, until the resulting fiber will not pass through the nucleus.
If this is not the case for any local stress maximum, the algorithm will go back to placing a random fiber instead.
The random fiber placement, in turn, generates random fibers as outlined above, until the created fiber will not pass through the nucleus. Table \ref{tab:gen-opt-parameters} summarizes all parameters used in the elastic model and the genetic algorithm.

\begin{table}[t]
    \centering
    \caption{Parameters used in the genetic algorithm.}
    \begin{tabular}{l r S[table-format=3.1,table-text-alignment=center] l}
        \toprule
        Topic & Quantity & {Value} & Unit \\
        \midrule
        Cell bulk medium (cytoplasm) & Young's modulus & 100 & \unit{\pascal} \\
         & Poisson ratio & {1/3} & \\
         & Isotropic prestress & 30 & \unit{\pascal} \\
        \midrule
        Stress fibers & Young's modulus & 10 & \unit{\kilo\pascal} \\
         & Tangential prestress & 1.2 & \unit{\kilo\pascal} \\
         & Initial mean radius & 50 & \unit{\nano\meter} \\
         & standard deviation (SD) of radius&  15 & \unit{\nano\meter} \\
        \midrule
        Mutation parameters & Fiber removal prob. $p_r$ & 0.7 & \\
        & Fiber variation prob. $p_v$ & 0.7 & \\
         & SD of fiber start/end translation & 400 & \unit{\nano\meter} \\
         & Fiber addition prob. $p_a$ & 0.7 & \\
         & Prob. for stress-aligned fiber placement $p_s$ & 0.8 & \\
         \midrule
        Nucleus bulk medium & Young's modulus & 1 & \unit{\kilo\pascal} \\
         & Poisson ratio & {1/3} & \\
         & Isotropic prestress & 0 & \unit{\pascal} \\
        \bottomrule
    \end{tabular}
    \label{tab:gen-opt-parameters}
\end{table}

\subsection{Optimal fiber distributions} \label{sec:optimal fiber distributions}

\begin{figure}[t!]
    \centering
    \includegraphics[width=\textwidth]{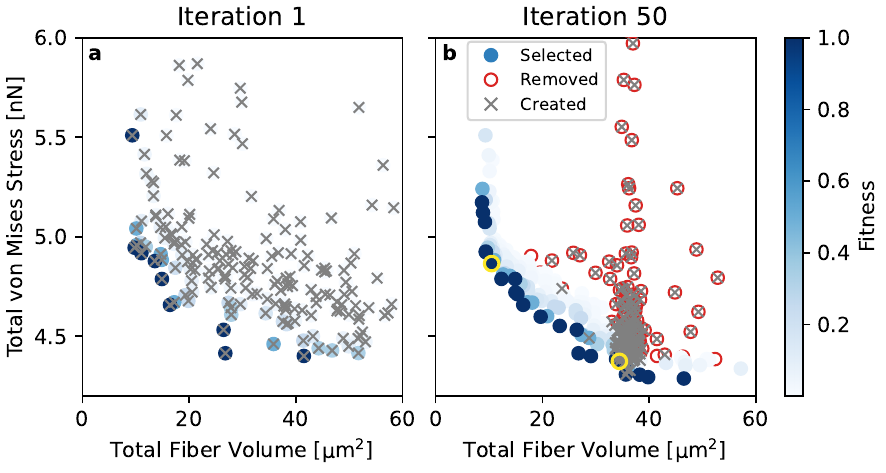}
    \caption{The objective function space with overall stress $z_1(\bm{x})$ 
and stress fiber volume $z_2(\bm{x})$
of the genetic algorithm with decision variables $\bm{x} \in R_t = P_t \cup Q_t$ for
(a) iteration $t=1$ and (b) iteration $t=50$ after selection was applied.
Blue markers denote selected decision variables $\bm{x} \in P_{t+1}$, with the color indicating the respective fitness value $f(\bm{x}; t+1)$.
Red circles denote variables that were not selected and hence removed from the population.
A cross additionally indicates variables $\bm{x} \in Q_t$ that were created in the respective iteration and thus are part of the most recent offspring population.
Note that in the first iteration no decision variables are removed, as no parent population exists yet.
The corresponding cell elasticity models of the two configurations marked by yellow outlines in (b) are displayed in Figure~\ref{fig:fiber-configurations}.}
    \label{fig:gen-opt-objective-functions}
\end{figure}

To demonstrate how the
optimization process works, Figure~\ref{fig:gen-opt-objective-functions} displays the objective function space for all decision variables $\bm{x} \in R_t = P_t + Q_t$ at the beginning (after just one iteration) and
after 50 iterations of the genetic algorithm.
We find that the overall stress, indicated by $z_1(\bm{x})$ on the ordinate, 
generally decreases with increasing stress fiber volume, $z_2(\bm{x})$ on the abscissa.
The Pareto front is convex in objective function space, and its sampling improves with the algorithm iterations,
cf.~Figure~\ref{fig:gen-opt-objective-functions}(a) vs.~(b).
While the initial spread is large, 
after several iterations
most decision variables lie close to the non-dominated front.
As can be seen by the coloring in  Figure~\ref{fig:gen-opt-objective-functions}(b), the algorithm usually creates an offspring population that is localized in objective function space, and discards most of these decision variables immediately (cf.~the red circles).

\begin{figure}[t!]
    \centering
    \includegraphics[width=\textwidth]{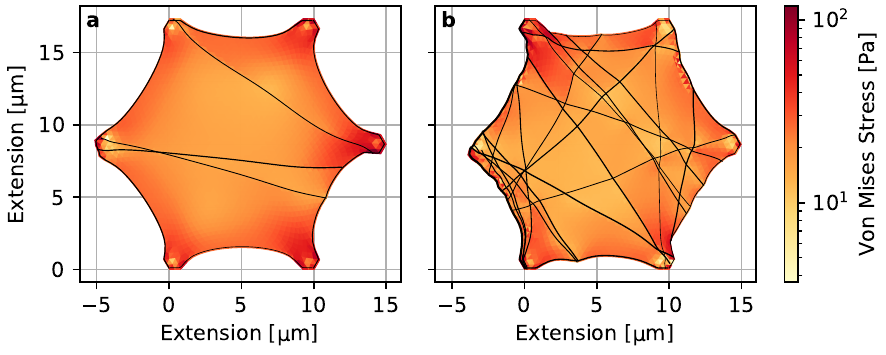}
    \caption{Two exemplary fiber configurations in the non-dominated front of the genetic algorithm after 50 iterations. Fibers are drawn as black lines whose line width indicates the fiber radius (not drawn to scale).
    Fiber radii in both configurations range from $\qtyrange{18}{113}{\nano\meter}$. (a)
    Low fiber material consumption
    ($z_2 =\qty{10.49}{\micro\meter\squared}$)
    but relatively large stress ($z_1 = \qty{4.87}{}$nN), also visible by the greater displacement to the left of the pulled corner.
    (b) Reduced stress ($z_1 = \qty{4.37}{}$nN) at the cost of using a lot more fiber material ($z_2 = \qty{34.43}{\micro\meter\squared}$).}
    \label{fig:fiber-configurations}
\end{figure}

Figure~\ref{fig:fiber-configurations} displays results for the cell elasticity model for two decision variables from the non-dominated front shown in Figure~\ref{fig:gen-opt-objective-functions}(b) (denoted there by the yellow circles).
They visualize the (near) extrema of the non-dominated front by optimizing the configuration for either (a) large stress and low total fiber volume or (b) low stress with large total fiber volume.
The configuration shown in (b) reaches an overall stress of $\lVert \sigma_\text{vM} \rVert$ = \qty{4.37}nN, compared to \qty{4.87}nN for configuration (a).
However, the total fiber volume required by configuration (b) is more than three times larger than that of (a), 
with $z_2 = \qty{34.43}{\micro\meter\squared}$ (b) against \qty{10.49}{\micro\meter\squared} (a).

\begin{figure}[t!]
    \centering
    \includegraphics[width=\textwidth]{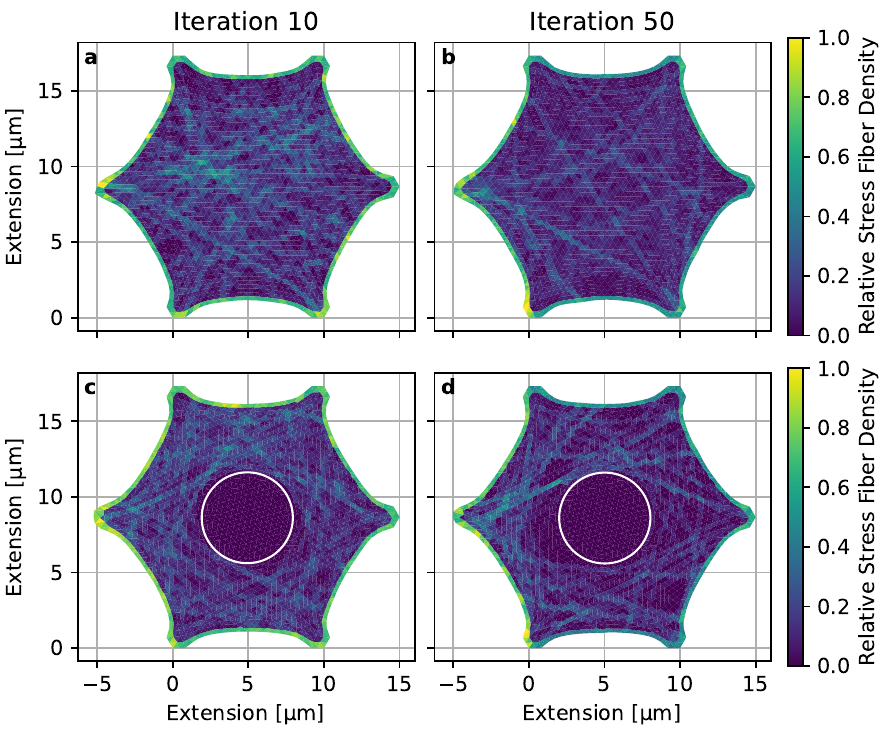}
    \caption{Relative stress fiber density averaged over all configurations with a fitness of one. 
    The geometry is given by the mean displacement of these configurations.
    (a) After 10 and (b) after 50 iterations of the genetic algorithm.
    (c) and (d) show the same simulations as (a) and (b), respectively, but now in the presence of an initially round nucleus in the cell center (outline indicated by white circle), through which stress fibers are not allowed to pass.}
    \label{fig:fiber-density}
\end{figure}

To evaluate and visualize also the resulting whole population of stress fiber configurations, we select the configurations with a fitness of one and call them population $O_t$, which typically is here of the size of around 10$\%$ of the whole population. We investigate the relative stress fiber density per grid element $T$, given by
\begin{equation}
    \rho_T = \frac{1}{\max_T \rho_T} \sum_{\bm{x} \in O_t} \sum_{i \in \mathcal{I}(\bm{x}, T)} 2 r_{\bm{x}, i},
    \label{rel_sf_density}
\end{equation}
where $\mathcal{I}(\bm{x}, T)$ collects the indices of fibers in $\bm{x}$ passing through grid element $T$, and $r_{\bm{x}, i}$ is the radius of the fiber with index $i$ in $\bm{x}$.

This relative stress fiber density is visualized in Figure~\ref{fig:fiber-density} 
over the mean displaced geometry of all $\bm{x} \in O_t$
for iteration (a) $t=10$ and (b) $t=50$. 
One can clearly see that the density distribution is non-symmetric, 
reflecting the effects of the force
applied to the left,
although the initial geometry is symmetric.
The highest stress fiber density is found at the peripheral arcs of the left side.
Stress fibers running through the cell tend to have at least one of their start or end points near an attached corner of the cell.
Fibers also appear to form nearly parallel groups.
After 50 iterations, we find that diagonal fibers, similar to the \emph{V} configuration in Figure~\ref{fig:growing}, are more prevalent than single horizontal fiber in the center that would directly oppose the exerted force.

To be even closer to the experimental situation, we study the same setup
including the cell's nucleus, as explained in the previous section.
Figure~\ref{fig:fiber-density}(c) and (d)
display the relative stress fiber density,
for iteration $t=10$ and $t=50$,
to be directly comparable to (a) and (b) in the absence of the nucleus.
A \emph{V} configuration 
starts to form after 10 iterations and is clearly visible after 50 iterations, as anticipated by the exclusion effect of the nucleus. 
As already observed without a nucleus, the peripheral fibers at the left side are strongly enhanced compared to the other four. Together with vertical fibers connecting the two attached corners left to the nucleus, these peripheral fibers form the \emph{Tri} configuration, the importance of which we already found in Figure~\ref{fig:extending} and which is also observed in experiments, cf.~Figure~\ref{fig:scheiwe}(f).
Overall, at $t=50$, the entire population shows a broken symmetry similar to the single configuration in the Figure~\ref{fig:fiber-configurations}(b), reinforcing the left side.

\begin{figure}[t!]
    \centering
    \includegraphics[width=\textwidth]{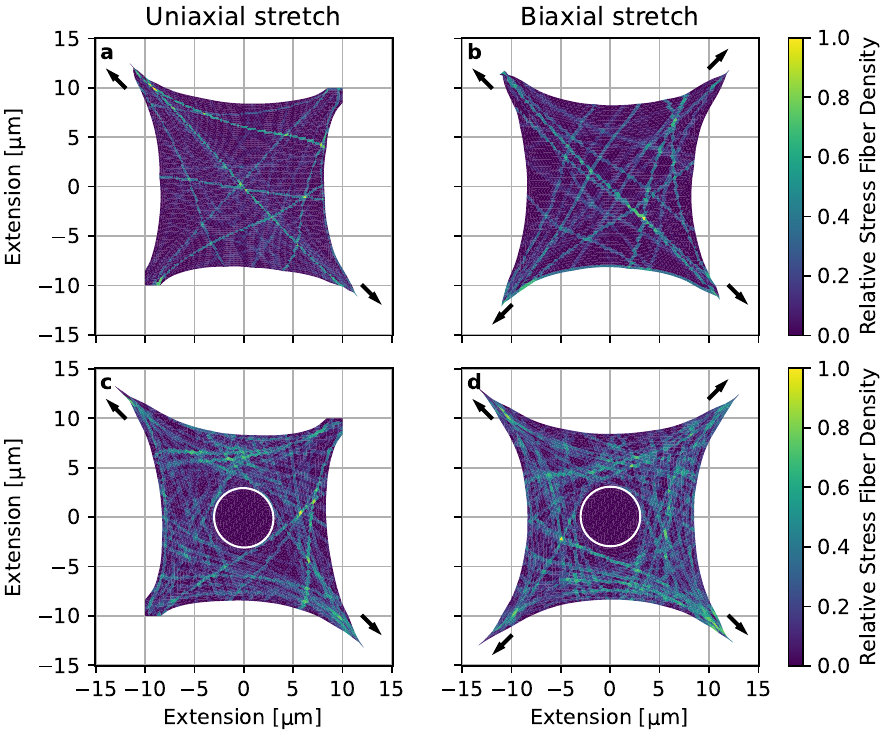}
    \caption{
    Relative stress fiber density averaged over all configurations with a fitness of one after 50 iteration of the genetic algorithm for stretching a square-shaped cell, where the arrows indicate the direction of applied force. The average is taken over all configurations with a fitness of one and the geometry is given by the mean displacement of these configurations.
    (a) Uniaxial and (b) biaxial stretch without a nucleus.
    (c) Uniaxial and (d) biaxial stretch in the presence of an initially round nucleus (outline indicated by white circle) in the cell center, through which stress fibers are not allowed to pass.
    }
    \label{fig:fiber-density-stretching}
\end{figure}

To demonstrate that our optimization procedure is generally applicable to different geometries and external force conditions, we also investigate uni- and biaxial stretching of a quadratic cell. We use the exact same values of the mechanical and optimization parameters as before (cf.~Table \ref{tab:gen-opt-parameters}). The resulting fiber distributions are shown in
Figure~\ref{fig:fiber-density-stretching}, in (a) and (b) without a nucleus and in (c) and (d) with a nucleus, respectively.  Without a nucleus, fibers form predominantly along the directions of the applied force (indicated by black arrows) and run from one to the opposite corner, as one would expect and has been observed experimentally for cells on substrates under static stretch \cite{katsumi_effects_2002}. Note that our discrete
approach allows for solutions in which 
the stress fibers cross each other. 
This stands in marked contrast to 
continuum descriptions of local actin orientation, 
which would predict vanishing orientation
at the crossing points \cite{deshpande_bio-chemo-mechanical_2006, farsad_xfem-based_2012, schakenraad_mechanical_2020}. In the uniaxial case, almost all other prominent fibers are anchored at least at one end to a Dirichlet boundary condition, i.e. to a focal adhesion, forming \emph{V}-~shaped configurations, which we have already observed in the previous hexagonal setup. These \emph{V} configurations are the main feature in the presence of a nucleus (Figure~\ref{fig:fiber-density-stretching}(c) and (d)), since then fibers cannot run between opposite corners anymore. The overall stress fiber distribution changes drastically, leading to a slightly less efficient reinforcement such that a little bit more fiber material is needed to reduce stress by the same amount. This is also reflected in the larger average displacement in Figure~\ref{fig:fiber-density-stretching}(b) and (d) compared to (a) and (c), respectively. Thus, also in reals cells the nucleus might present an important geometrical challenge depending on the geometry of the attachment points and focal adhesions. However, most of the fibers are still anchored to one of the stretched corners and are able to reduce displacement and therefore stress substantially compared to a situation without any fibers. In \ref{sec:sheared cells}, the predicted fiber distribution of a sheared cell is shown as another example.

\section{Discussion}

In this work, we have presented a novel approach to simulate the mechanics of biological cells with contractile actin stress fibers.
We modeled the fibers as prestressed, one-dimensional \acf{eb} beams embedded in a bulk elastic medium using the CutFEM method developed by \cite{hansbo_cut_2017}. Our simulations of stress fiber positioning after mechanical stimulation
show very good agreement with the experimental observations reported by \cite{scheiwe_subcellular_2015},
demonstrating that these configurations minimize 
internal mechanical stress. As a measure
for this stress, we used the $L^1$-norm of the von Mises stress in the bulk. 
The optimization procedure was implemented
with genetic algorithms, minimizing 
both the stress and the amount of actin used for stress fibers in the cell. Our work demonstrates that typical stress fiber structures within cells can be predicted by focusing on mechanics
and not considering biochemical details.

A central element of our work is the requirement to efficiently simulate the effect of stress fibers. 
Here we use the recently introduced method of CutFEM 
\cite{hansbo_cut_2017} for embedded fibers.
Our simulations in  section~\ref{sec:fiber-compare}
demonstrate that modelling cellular stress fibers as EB beams with prestress via CutFEM
is  in very good agreement with simulating the fibers explicitly.
Only slight differences are visible near the cell boundary and at the end of fibers due to the singular character of the 1D fibers, see Figure~\ref{fig:comparison}. Employing finer meshes or local mesh refinement around the ends of the fibers reduces these effects.

The main advantage of the \ac{eb} model is that the fiber configuration is independent of the bulk material grid.
Especially when many fibers are simulated, the generation of a grid that needs to explicitly resolve all fibers can be tedious or must be automated.
Even so, the number of grid nodes, and hence the computational cost, rises strongly with the number of fibers because their diameters (\qty{100}{\nano\meter}) are small compared to the extensions of a cell
(\qty{10}{\micro\meter}).
The \ac{eb} model therefore enabled us to compute solutions for numerous stress fiber configurations based on very few grid files, and to automate this process when using the genetic algorithm.
In the future, it would be very interesting to extend this procedure from 2D to 3D situations,
especially for experimental situations in which cells are not in the effectively 2D situation shown in Figure~\ref{fig:scheiwe}.
As a downside, the discontinuous Galerkin formulation of the model by \cite{hansbo_cut_2017} is difficult to implement, mostly due to the geometric transformations required to compute the normal and tangential derivatives.

Linear elasticity as used here is only a first-order
approximation for the complex mechanics of cells
\cite{pullarkat2007rheological,kollmannsberger_linear_2011}.
A more detailed approach might consider
the non-linear elasticity that emerges
for networks of semiflexible polymer networks as the cytoskeleton \cite{broedersz2014modeling}.
In principle, our approach can be
extended to nonlinear elasticity in the future, as the bulk elasticity and the fiber model couple only through the common displacement field.
This fact in particular enables the presented linear \ac{eb} model to be combined with a non-linear model for bulk elasticity.

Our approach also simplifies several features of actin stress fibers.
In adherent cells, stress fibers are often anchored by focal adhesions, which connect them to the extracellular matrix \cite{tojkander_actin_2012}.
It is less clear how their ends are
organized if not anchored to adhesions.
In our model, fibers are allowed to end at any point inside the cell.
As special anchoring points are not considered, the fiber (pre)stress must then be compensated by the elastic medium surrounding the fiber end.
Since the Young's modulus of the fiber exceeds the one of the bulk medium by several orders of magnitude, this can lead to non-physical solutions where the bulk medium folds in on itself, an effect
that could be avoided by implementing
explicit anchoring points.
We also assume that a bulk elasticity model based on finite strain theory might alleviate this issue.

For simplicity, we considered only homogeneous fibers, i.e.~with constant stiffness and radius.
It has been shown in \cite{lu_mechanical_2008} that stress fibers can appear heterogeneous towards strong contractile levels, and mechanically non-linear towards weak contractile levels.
While heterogeneity is conceptually included 
in \ac{eb} beam theory
and could be directly studied with the approach reported here, beam nonlinearity is not and would require additional modeling efforts.

Our simulations in section~\ref{sec:fiber-growth} demonstrate that fiber configurations observed in experiments reduce the stress in the cell, both individually and jointly.
We have found the global von Mises stress $\lVert \sigma_\text{vM} \rVert$ to be a good quantification of the overall cell stress, as
it decreases monotonically for increasing fiber lengths and reaches a minimum when all fiber configurations found in experiments are combined.
In contrast, the global strain energy $\lVert \mathcal{W} \rVert$ is not strictly decreasing with fiber length and therefore less suitable. In the future, one could identify
other measures of internal cell stress that work well, including contractions of higher order tensors representing network organization.

The genetic algorithm of section~\ref{sec:fiber-cfgs} produces stress fiber configurations that resemble those found in experimental studies.
This demonstrates that it is feasible to define the problem of finding optimal fiber configurations as a multi-objective optimization task, where both the stress in the cell and the amount of actin used to build stress fibers are sought to be minimized.
Our approach ignores underlying biochemical mechanisms and operates only on a cell-macroscopic mechanical level.
Our results serve as a proof of concept, indicating that modeling biochemical mechanisms within the cell are not necessarily required when investigating the formation and configuration of stress fibers, especially when focusing on their mechanical aspects.

Testing our algorithm with the same optimization parameters for a different cell geometry for uni- and biaxial stretching as well as for shear forces (see \ref{sec:sheared cells}) again resulted in
realistic fiber configurations. In particular,
our algorithm predicts that discrete stress
fibers can cross each other, 
which is a remarkable result that
cannot occur in a continuum theory. 
This validates our hypothesis of global stress minimization and demonstrates the applicability of the optimization procedure without additional adaptation to the specific problem. 
A nucleus representing an excluded volume, which cannot be intersected by the fibers, can change the resulting fiber distribution substantially.
Depending on the geometry of the investigated problems, it might hinder the formation of very effective fibers, like in the case of uni- and biaxial stretching, or it singles out a very distinct fiber configuration from a former more homogeneous distribution.
In the future, 
our purely mechanical approach could be complemented by biochemical aspects, including signaling through the Rho-system, which is the main regulator of contractile actin architectures \cite{banerjee_actin_2020,staddon2022pulsatile,andersen_cell_2023}.

Our approach is mainly phenomenological, as it focuses on few global parameters in the mechanical system of the cell.
It is clear that biological cells do not actively solve an optimization problem like the one we implemented.
They have a limited actin pool to build stress fibers from and are unable to sense a global quantity like the von Mises stress $\lVert \sigma_\text{vM} \rVert$.
However, they are able to sense local stress via mechanosensing pathways that trigger actin condensation and subsequently form actin strands in the direction of principal stress.
As the actin filaments have a much higher stiffness than the cellular cytoplasm, the filament formation reduces stress in the cytoplasm.
Active stress fiber contraction then also decreases strain if it is able to counteract the force exerted on the cell.
These mechanisms contribute to a decrease in stress and strain throughout the cell and we therefore deem $\lVert \sigma_\text{vM} \rVert$ a sensible global measure albeit originating from a local
biochemical-mechanical interplay.

Our study reported here is not exhaustive and the optimization process could be further improved to possibly yield results that are more consistent and closer to reality.
Our current implementation performed optimization by changing fiber positions and fiber radii.
Other fiber parameters like Young's modulus and prestress could also be added to the genetic algorithm.
In principle, one would expect stiffer fibers without prestress could oppose an external stretching force equally well as a softer, prestressed fiber. However, to reduce the overall displacement about the same amount under the same conditions, the stiffness would have to be increased more than twofold in our simulations, which would not be in the physiological range of real stress fibers. Furthermore, prestressed fibers in the wrong place might increase displacement of the bulk unnecessarily and increase stress drastically if their ends are not anchored to focal adhesions. Therefore, prestress facilitates the efficient placing of fibers. Without fiber prestress, the resulting fiber distribution obtained with our genetic optimization qualitatively differs from the here shown results in Figure~\ref{fig:fiber-density} with prestress and reproduces the experimental images less well.
If verified experimentally, predictions for these parameters would provide valuable information because assessing fiber parameters \emph{in vitro} is still difficult.
Also note that
due to a small total fiber number -- which however is realistic for cells --  the results produced by the current genetic algorithm depend on the generation of random numbers.
Fiber configurations can vary strongly between different runs with the same parameters but different seeds.
Most resulting populations are dominated by few fiber configurations, which might be remedied by a higher population size.
However, even though a single run of the genetic algorithm does not necessarily explore all fiber configurations found in experiments, we find that it samples the objective function space well, cf.~Figure~\ref{fig:gen-opt-objective-functions}. Therefore, the used population size is sufficient to obtain good agreement with the averaged experimental fiber distribution.
The genetic algorithm has a low acceptance rate, meaning that only few configurations from offspring populations are actually selected in the next iteration.
This is the reason why we chose an elitist algorithm in the first place.
Single point crossover between Pareto-optimal configurations seems to produce mostly inferior configurations in our case.
We conclude that the underlying assumption of the genetic algorithm, namely that fibers in a decision variable are independent, may not be truly justified.
In principle this is intuitive:
Any fiber that does not run along the symmetry axis of the problem (in the hexagonal setup, the horizontal) needs a symmetric counterpart to avoid deformation into its direction.
Our implementation so far neglects these relations between fibers within a configuration.
A significant improvement for a fiber configuration search algorithm would be to consider such beneficial correlations between fibers.
It could then perform an optimization based on structures consisting of multiple fibers like the ones we investigated in section~\ref{sec:fiber-growth}.
Presumably, a machine-learning algorithm would be better suited for such a task
than a genetic algorithm.

Finally, as we have formulated an optimization problem based mostly on fiber positioning, transient aspects of fiber creation are not considered in this approach.
As actin polymerization and myosin II contractility in cells is locally initiated in regions in which stress is high, the fiber distribution in real cells necessarily depends on the dynamics of this process. It would be interesting
to also study these aspects of fiber growth
in the future.

\section*{Data accessibility} 

Our computer code is publicly available at \url{https://github.com/usschwarz/dune-structures}.

\section*{Authors’ contributions}

All authors contributed to the conceptualization of this work. 
LR, VW and DK performed the work and implemented the code. 
All authors analyzed and discussed the simulation results.
PB and USS supervised the projected and acquired funding.
LR and USS wrote the original draft. 
All authors reviewed and commented on the text.

\section*{Conflict of interest declaration} 

We declare that we have no competing interests. 

 \section*{Funding} 

This work is funded by Deutsche Forschungsgemeinschaft (DFG, German Research Foundation) under Germany’s Excellence 
Strategy - EXC 2181/1 – 390900948 (the Heidelberg STRUCTURES Excellence Cluster).

\section*{Appendix: Optimal fiber distribution of a sheared cell} \label{sec:sheared cells}

\begin{figure}[t!]
    \centering
    \includegraphics[width=\textwidth]{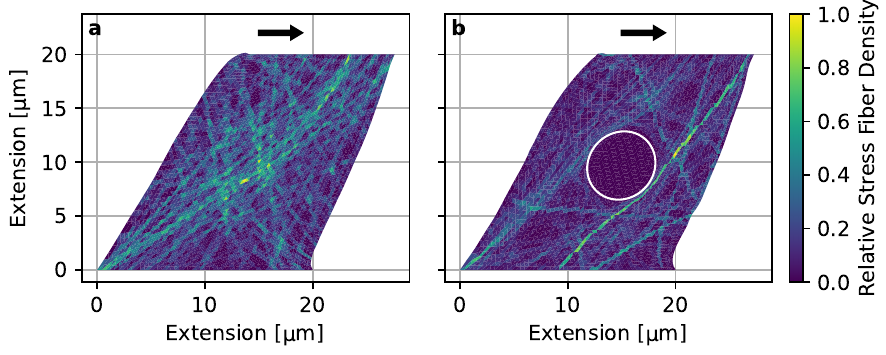}
    \caption{
    Relative stress fiber density averaged over all configurations with a fitness of one after 40 iteration of the genetic algorithm for shearing a square-shaped cell, where the arrows indicate the direction of applied force, while the cell's position is fixed at the bottom as well as the height in vertical direction. The average is taken over all configurations with a fitness of one and the geometry is given by the mean displacement of these configurations.
    (a) Without a nucleus and (b) in the presence of an initially round nucleus (outline indicated by white circle) in the cell center, through which stress fibers are not allowed to pass.
    }
    \label{fig:fiber-density-shear}
\end{figure}

To test our algorithm in a slightly different, less structured setting, we consider a initially square-shaped cell, which is attached to a substrate along the whole length of one edge, instead of discrete anchoring points investigated in section~\ref{sec:optimal fiber distributions}. When then apply a force, parallel to the substrate, at the top of the cell to shear it, while keeping its height fixed. In figure~\ref{fig:fiber-density-shear} the resulting fiber distribution of our genetic algorithm is shown after 40 iterations. In (a), without a nucleus, fibers mainly form from the lower left to the upper right corner to oppose the external force, while being anchored to the substrate. In general, fibers tend to run from the bottom to the top, as one would expect. In the presence of a nucleus, which cannot be crossed by the fibers, two main fibers, running vertically on the left and right side of the nucleus, emerge. As observed before, the nucleus hinders the most efficient fiber formation and more fiber material is needed for the same stress reduction.

Even though these setup is very different compared to the ones analyzed in section~\ref{sec:optimal fiber distributions}, the algorithm still converges within the same number of iterations and produces meaningful fiber distributions without adjusting the values of the optimization parameters.


\end{document}